\documentclass[prl,aps,twocolumn,preprintnumbers,showpacs,superscriptaddress]{revtex4}

\usepackage{graphics,graphicx,epsfig,bm,amsthm,amsmath}

\newcommand{\ket}[1]{\left| #1\right\rangle}      

\begin{document}

\title{Simulation of Classical Thermal States on a Quantum Computer: A Transfer Matrix Approach}

\date{\today}

\author{Man-Hong Yung}
\email{mhyung@chemistry.harvard.edu}
\affiliation{Department of Physics, University of Illinois at
Urbana-Champaign, Urbana IL 61801-3080, USA}

\affiliation{Department of Chemistry and Chemical Biology, Harvard University, Cambridge MA, USA}

\author{Daniel Nagaj}
\affiliation{Research Center for Quantum Information, Institute of Physics,Slovak Academy of Sciences, D\'{u}bravsk\'{a} cesta 9, 845 11 Bratislava, Slovakia}

\author{James D. Whitfield}
\affiliation{Department of Chemistry and Chemical Biology, Harvard University, Cambridge MA, USA}

\author{Al\'{a}n Aspuru-Guzik}
\email{aspuru@chemistry.harvard.edu}
\affiliation{Department of Chemistry and Chemical Biology, Harvard University, Cambridge MA, USA}


\begin{abstract}
We present a hybrid quantum-classical algorithm to simulate thermal states of classical Hamiltonians on a quantum computer. Our scheme employs a sequence of locally controlled rotations, building up the desired state by adding qubits one at a time. We identify a class of classical models for which our method is efficient and avoids potential exponential overheads encountered by Grover-like or quantum Metropolis schemes. Our algorithm also gives an exponential advantage for 2D Ising models with magnetic field on a square lattice, compared with the previously known Zalka's algorithm. 
\end{abstract}

\maketitle
\begin{figure}[t]
\includegraphics[bb=10 20 450 300,scale=0.5]{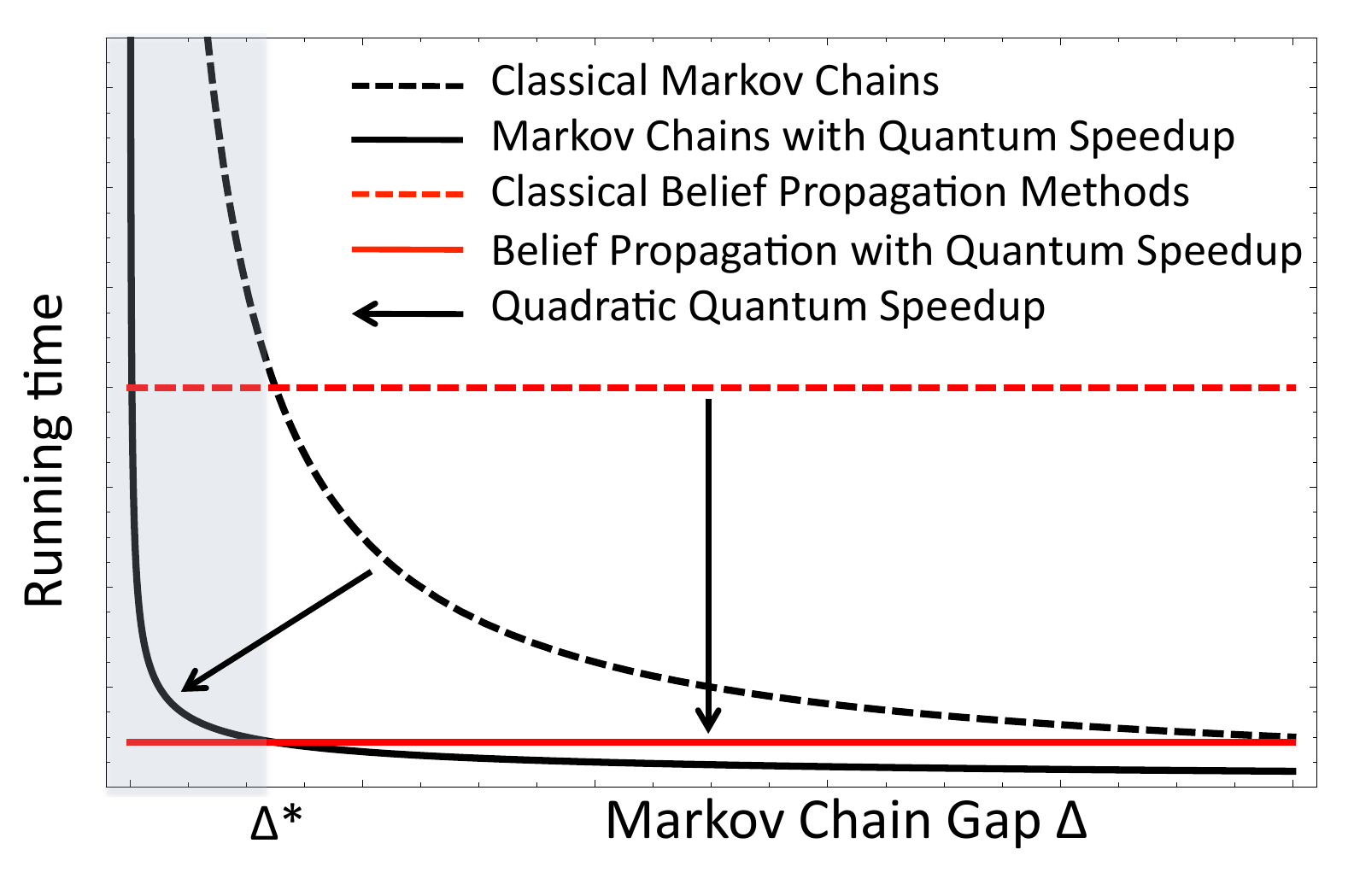} 
\caption{(color online) The running time ${\tau \sim O\left( {1/\Delta } \right)}$ of Markov chain methods is limited by the gap $\Delta$ of the Markov matrix. A quantum quadratic speedup ${\tau \sim O( {1/\sqrt \Delta  } )}$ (black solid line) relative to classical Markov chains (black dashed line) can be achieved by a quantum computer. Below some critical gap size ${\Delta<\Delta ^*}$, Markov chain methods become inefficient (shaded region), and classical belief propagation methods (red dashed line), including transfer matrix methods, become more efficient. Combined with amplitude amplification, a further quantum speedup is possible (red solid line).}\label{fig:MC_scale}
\end{figure}

Simulation of a finite-temperature physical system with a controllable quantum device is one of the most important goals of quantum simulation~\cite{Buluta2009,Kassal2010}. Classical Markov-Chain Monte Carlo (MCMC) algorithms are powerful tools for sampling Gibbs distributions. They are efficient provided that the gap $\Delta$ of the transition matrix is non-vanishing; the running time typically scales as $\tau \sim O\left( {1/\Delta } \right)$. A quantum generalization  \cite{Szegedy04} of MCMC has recently been explored by the quantum information community due to the connection to quantum walks \cite{Aharonov2001}. Richter \cite{Richter07} developed a method for sampling from the Gibbs distribution for periodic lattices. Somma {\it et al.} \cite{Somma2008} combined quantum walk and quantum Zeno effect to achieve quantum speedup. Wocjan and Abeyesinghe \cite{Wocjan08} improved it by using fixed point quantum search. Generally, these quantum algorithms allow the running time to scale as $\tau \sim O( {1/\sqrt \Delta  })$, a quadratic speedup compared with the classical counterparts. However, for many problems of practical interest, such as optimization problems and spin glasses, the gap $\Delta$ may become exponentially small when the system size increases, making it unpractical to use MCMC algorithms for solving them (see Fig. \ref{fig:MC_scale}). Therefore, gap-independent methods are more desirable for solving these problems. 

A class of gap-independent methods is called \emph{belief propagation} \cite{BP}, which generalizes the transfer matrix methods in statistical physics. For problems involving a regular geometry, it can be very efficient, e.g. one-dimensional spin chains. This property will be exploited in this letter, where a different way for obtaining samples from the thermal state is discussed. This approach is a generalization of the state preparation method by Lidar and Biham \cite{Lidar97}, and Zalka \cite{Zalka98}. We show that in some cases, the structure of the system under investigation allows for large speedups over the general methods.  This is because the cost of our method is independent of the temperature and the gap size. 

Our proposed strategy is to construct a \emph{coherent encodings of a thermal state} (CETS) $\ket{\psi_{CETS}}$ directly, rather than sampling from the thermal probability distribution:
\begin{equation}
\left| {\psi _{CETS} } \right\rangle  = \sum\limits_s {\sqrt {e^{ - \beta H\left( s \right)} /Z} } \left| s \right\rangle,
\label{CETS}
\end{equation}
where $s = \{ {0,1} \}^N$, $\beta  = 1/{k_B}T$ is the inverse temperature, $H(s)$ is the eigen-energy of some classical spin Hamiltonian for the $N$-spin configuration $s=s_1 s_2 \cdots s_N$ and $Z$ is the partition function.
This CETS can be transformed into the corresponding thermal state $\rho  = {e^{ - \beta H}}/\textrm{Tr}\left( {{e^{ - \beta H}}} \right)$ by including a set of $N$ ancilla qubits, 
performing bit-by-bit CNOTs such that 
$\ket{s}\otimes\ket{0\cdots0}_A \to \ket{s} \otimes \ket{s}_A$ 
and tracing over the ancilla system.
However, for some applications, such as the partition functions estimation in \cite{Wpartition}
it is preferable to use the CETS directly.

Below we present a method for preparing the CETS of a classical Hamiltonian from 
the initial state $\ket{0\cdots0}$ by a sequence of locally-controlled rotations.
Zalka's approach \cite{Zalka98}, as applied to discrete cases \cite{Kaye2004}, allows for preparing the CETS
by adding qubits one by one, and performing a rotation (controlled by \emph{all} of the previous qubits) 
on each new qubit as
\begin{equation}
\ket{s_1 \cdots s_k} \ket{0}  \to 
\ket{s_1 \cdots s_k } \left( \cos \theta_s \ket{0}  + \sin \theta_s \ket{1} \right), \label{controlROT}
\end{equation}
where ${\cos^2 \theta_s }$ is the conditional probability of $s_{k+1}=0$, given that the first $k$ spins are in a particular configuration $s_1 s_2 \cdots s_k$. The problem here is 
that in general, this requires the knowledge (or efficient calculation) of $O\left(2^N\right)$ conditional probabilities. Thus, Zalka's method is efficient only when the probability distributions are efficiently integrable \cite{Grover2002}. Here we focus on the cases where the controlled rotations are local, i.e., they depend only on a few previous qubits. This in turn allows efficient computation of the respective rotation angles. More precsiely, we take into account the {\it structure} of the geometry of the physical systems
and use the idea of the renormalization approach to obtain the rotation angles.

\begin{figure}[t]
\includegraphics[bb=50 20 350 370,scale=0.5]{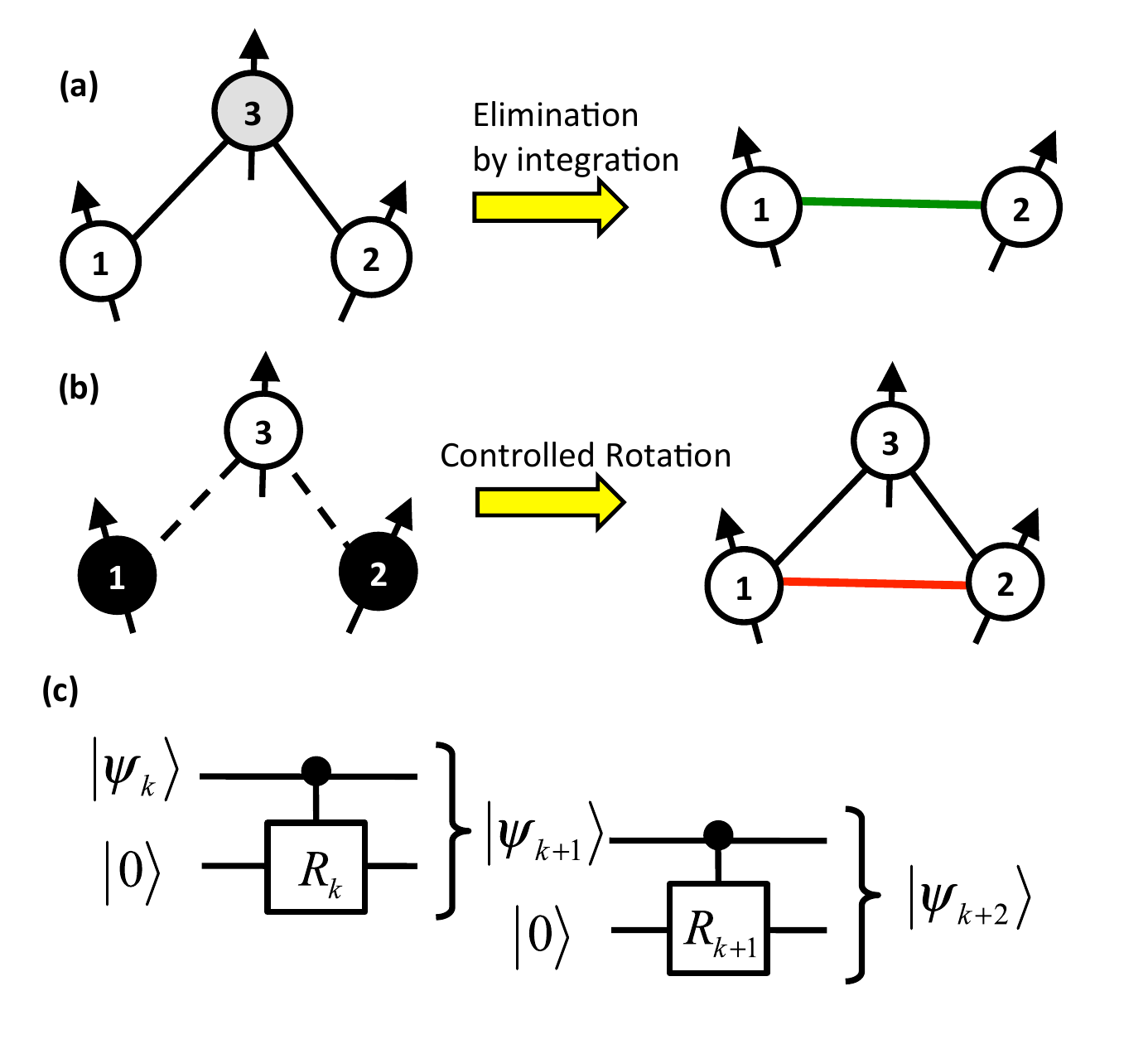} 
\caption{(color online) Real space renormalization approach for preparing the coherent encoding of a thermal state (CETS). (a) Spin 3 is eliminated by integration, inducing an effective interaction (blue bond) between spin 1 and spin 2. (b) A controlled rotation is performed on spin 3, inducing an effective interaction (red bond) between spin 1 and spin 2. (c) A quantum circuit demonstrating the sequential construction of the full thermal state.}\label{fig:RG}
\end{figure}

\emph{Real-space renormalization ---}
This method is also related to the renormalization group method \cite{Grosso}, which idea is to integrate out some degrees of freedom (coarse-graining) in the partition function $Z$, and describe the sub-system with a similar system with modified (renormalized) couplings. As an example, consider a linear chain of three spins (Fig {\ref{fig:RG}}a). The partition function after eliminating spin 3 (cf. Eq.(\ref{renorm_cond_1})),
\begin{equation}
Z = \Lambda \left( \beta  \right)\sum_{s_1 ,s_2 } {e^{B\left( \beta  \right)s_1 s_2}}
\end{equation}
is proportional to that of spin 1 and spin 2 interacting with an effective interaction $-B\left( \beta  \right)/\beta$. In contrast to this conventional renormalization treatment, where the degrees of freedom of the physical systems are progressively reduced, our method works in a reverse fashion: at each step, we \emph{increase} the number of degrees of freedom, and then perform a controlled-rotation (Fig. {\ref{fig:RG}}b), which also changes the effective interaction of spins 1 and 2.

Below, we define a sequential method for preparing a CETS for a generalized Ising Hamiltonian of $N$ classical spins which has multiple spin-spin coupling constants:
\begin{equation}\label{general_H_n}
	H_s = \sum\limits_j {A_j}{s_j} 
		+ \sum\limits_{ij} {B_{ij}}{s_i}{s_j} 
		+ \sum\limits_{ijk} {C_{ijk}}{s_i}{s_j}{s_k} + \dots
\end{equation}
Our goal is to investigate how a CETS can be constructed by locally-controlled quantum rotations. Suppose we are given a CETS as defined in Eq. \eqref{CETS} of $k$ spins $\ket{\psi_k}$ for the Hamiltonian given by Eq. (\ref{general_H_n}), and an additional qubit initialized in the state $\ket{0}$ that will become spin $k+1$ of our CETS. Let us define the rotation angle, $\theta_s$, by
\begin{equation}
	\cos {\theta _s} \equiv \sqrt {{e^{ - \beta {m_s}}}/W_s},  \label{cosine_rotation} 
\end{equation}
where $W_s \equiv e^{ - \beta m_s }  + e^{\beta m_s }  = 2\cosh \left( {\beta m_s } \right)$, and ${m_s} \equiv m\left( s_1,\dots,s_k \right)$ is a function of the spin variables of the first $k$ spins. 
After performing a controlled rotation \eqref{controlROT} on spin $k+1$, 
with angles given by Eq. \eqref{cosine_rotation}, we obtain a CETS $\ket{\psi_{k+1}}$ of a new $(k+1)$-spin Hamiltonian  
\begin{equation}
	H_{k + 1} = \tilde{H}_k + m_s \cdot s_{k + 1}. \label{renormalizedH}
\end{equation}
To justify this statement, rewrite $W$ in Eq. \eqref{cosine_rotation} as
\begin{eqnarray}
	W_s = e^{-\beta m_s} + e^{\beta m_s} = \Lambda_k e^{-\beta \left(H_k - \tilde{H}_k\right)}, \label{defineWs}
\end{eqnarray}
for some constant $\Lambda_k$ and some $k$-spin Ising spin Hamiltonian $\tilde{H}_k$ (with possible higher-order interactions). The state that we get from $\ket{\psi_k}\ket{0}$ by the controlled rotation \eqref{cosine_rotation} is $\sum_s {\sqrt {F(s)}}\ket{s}$ with $F(s)\equiv F(s_1,s_2,\dots,s_n)$ given by
\begin{equation}
	F(s)= \frac{e^{-\beta H_k}}{Z_k}  \times \frac{e^{-\beta m_s\cdot s_{k+1}}}{W_s} =\frac{e^{-\beta \left( \tilde{H}_k + m_s \cdot s_{k+1}\right)}}{Z_k \Lambda_k} 
\end{equation}
i.e. a CETS for the Hamiltonian \eqref{renormalizedH}. Moreover, 
the new normalization constant
is the same as the partition function ${Z_{k + 1}}$ for the system with Hamiltonian \eqref{renormalizedH} and can be obtained simply by 
$	{Z_{k + 1}} = {\Lambda_k}{Z_k}$.
The term ${{\tilde H}_k}$ in \eqref{renormalizedH} is an Ising Hamiltonian of the form \eqref{general_H_n}
for the first $k$ spins, but associated with a different set of renormalized couplings $\{ {{{\tilde A}_j},{{\tilde B}_{ij}},{{\tilde C}_{ijk}},...} \}$. 
Finally, the constant ${\Lambda _n}$ can be shown to be the geometric mean of the LHS of \eqref{defineWs}
\begin{equation}
	\Lambda_k  = 2\prod\limits_{m_s } \left[ \cosh \left( 2\beta m_s \right) \right]^{1/2^k}. 
\end{equation}
This is reminiscent of formulas which appear in classical algorithms such as belief propagation \cite{BP} for calculating some thermal properties of some spin systems. The controlled rotation is therefore the crucial element of our renormalization step. 
Using our method iteratively as shown in Fig. \ref{fig:RG}c, we can 
generate the CETS of a particular spin Hamiltonian. 

In general, $\tilde{H}_k$ could contain up to $k$-local interaction terms.
If all the terms in $\tilde{H}_k$ involve at most $t$ spins, we call this a $t$-\textit{renormalizable} operation. From now on, we will restrict ourselves to $2$-renormalizable operations, which for example, includes local magnetic fields and two-spin interactions. Next, we will consider a general construction of a CETS for systems with finite range interaction, followed by some specific examples for further illustration.

\emph{Finite-range interactions and belief propagation  ---} 
As a general construction, we consider spin chains with finite-range interactions involving $z$ neighboring spins. The computational complexity of this approach generally scales exponentially in $z$. As an example, consider two groups of spins $s$ and $t$, each can be considered as a $2^z$ dimensional system. The Hamiltonian is of the form
\begin{equation}
	H = H_s  + H_t  + H_{st},
\end{equation} 
where $H_s$ and $H_t$ are the internal interaction terms for spins within group $s$ and $t$, and $H_{st}$ contains the interactions between the groups. 
We start with preparing the state of the group $s$ as 
$M^{ - 1/2} \sum\nolimits_s {\sqrt {e^{ - \beta H_s }  \cdot \gamma _s } } \ket{s}$, where $M$ is a normalization constant, and $\gamma_s$ is a function of the spins $s$ to eliminate renormalization effects induced by the spins in group $t$. The group $t$ is initialized in the state $\ket{0\cdots 0}$.
We choose the controlled rotation \eqref{controlROT},
\begin{equation}\label{blockspin_c_rot}
\ket{s}\ket{0\dots 0} \to \ket{s} 
\sum\limits_t \sqrt {e^{ - \beta \left( {H_t  + H_{st} } \right)}  
										\cdot \left( {\gamma _t /\gamma _s } \right)}  \ket{t},
\end{equation}
with $\gamma _t$ determined by the next group of spins to be included in the preparation procedure. If group $t$ is the last group, then all $\gamma_t $ are equal to $1$. To ensure unitarity of this operation, we require 
\begin{equation}
	\gamma _s  = \sum\limits_t  \gamma _t e^{ - \beta \left( {H_t  + H_{st} } \right)} ,
\end{equation}
which is a recursion relation typically encountered in belief propagation\cite{BP} problems. For a group of $z$ spins, and a given set of $\gamma_t$, the sum involves $O(2^z)$ terms, scaling exponentially in $z$. To perform the multi-qubit rotation, we can apply Zalka's algorithm \cite{Zalka98}, which requires the computation of $O(2^z)$ rotation angles, and a polynomial number of subsequent quantum operations. To save computational resource for large $z$, it is more efficient to determine the angles for rotation ``on the fly". This can be achieved by the quantum amplitude amplification algorithm \cite{1206629} calculated with some ancilla qubits. 

We can apply this approach to an $N\times N$ square lattice of Ising spins with non-uniform couplings and arbitrary local magnetic fields. We make a group for each row of $z=N$ spins. In the worst case scenario, the number of required operations in the above approach then scales \footnote{With belief propagation, for a chain of $d$-dimensional qudits, the partition function scales as $(N-1) d^{2}$.} as $O(2^{2N})$, which becomes $O(2^{N})$ after combining with the amplitude amplification algorithm. This is still an exponential algorithm, but nevertheless with an exponential speed up over the direct application of Zalka's algorithm, whose complexity scales as $O(2^{N^2})$, as it requires the preparation of a probability distribution with $2^{N^2}$ amplitudes. 
However, for the uniform 2D Ising model without magnetic fields, an efficient $t$-renormalizable approach might exist, as classical polynomial algorithms exist for this problem \cite{Barahona82}.

In the following, we illustrate our method by explicitly giving several examples of physical interest.

\emph{Building blocks for frustrated magnets and spin ice ---}
As the first example, we show how to generate a CETS of a triangle plaquette of three Ising spins by a 2-renormalizable operation.
Our goal is to prepare a CETS of three spins (see Fig. \ref{fig:RG}b), for the Hamiltonian 
$H_3 = J{s_1}{s_2} + J{s_1}{s_3} + J{s_2}{s_3}$. 
Let us start with two qubits initialized as: 
$M^{-\frac{1}{2}} \sum\nolimits_{s_1, s_2 =\{0,1\}} 
\sqrt {\gamma _{s_1 s_2 } e^{ - \beta Js_1 s_2 } } \ket{s_1 s_2}$, where $M$ is a normalization constant and $\gamma _{s_1 s_2 }  > 0$ is some positive function of $s_1$ and $s_2$ to be determined later. Let us add a third qubit in the state $\ket{0}$ to the system, and act with the controlled rotation \eqref{cosine_rotation} depending on the values of the first two qubits. 
When we choose $m_s = J(s_1+s_2)$ for some constant $J$,
we can use the well-known result in renormalizing the 1D Ising chain \cite{Grosso}, and write
\begin{equation}\label{renorm_cond_1}
W_s = {e^{-\beta J\left( {{s_1} + {s_2}} \right)}} + {e^{ \beta J\left( {{s_1} + {s_2}} \right)}} 
= \Lambda {e^{\beta B{s_1}{s_2}}},
\end{equation} 
where the coefficients $\Lambda$ and $B$ are 
\begin{eqnarray}
	\Lambda &=& 2\sqrt {\cosh \left( {2\beta J} \right)}, \label{Lambda_and_B}\\
	B &=& (1/2\beta) \ln \cosh \left( {2\beta J} \right). \nonumber
\end{eqnarray}
Observe now that if we chose $\gamma _{s_1 s_2 } = \Lambda e^{  \beta Bs_1 s_2 }$ when preparing the first two qubits, applying the controlled rotation of the third qubit eliminates this factor. Consequently, this operation produces the CETS for the 3-spin Ising cycle $H_3$. 

\begin{figure}[t]
\includegraphics[bb=10 10 600 520,scale=0.3]{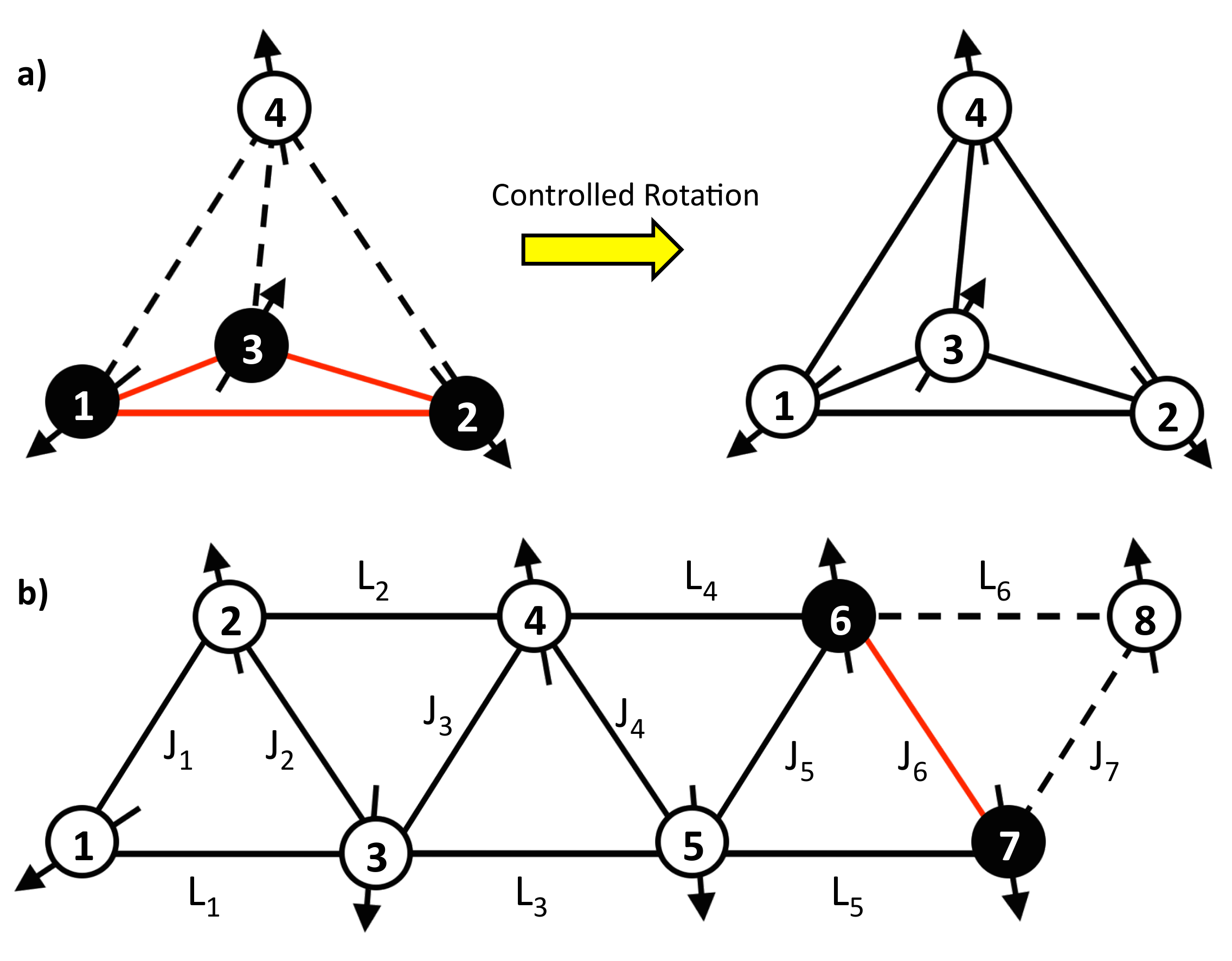} 
\caption{(color online) Examples of thermal state generation for the basic units of Ising spin cycles and spin ice (a) A rotation on spin 4, controlled by spin 1,2, and 3, results a thermal state of the tetrahedral spin ice.  (b). A next-nearest-neighbor spin chain is created by consecutive application of two-qubit controlled rotations.}\label{fig:ice}
\end{figure}

Furthermore, we can build on top of this triangular lattice to prepare a CETS of the basic tetrahedral unit of spin ice \cite{Bramwell2001} (see Fig. \ref{fig:ice}(a)).  Let's add a fourth qubit to the system, and perform a rotation controlled by the first three qubits:
\begin{equation}
\left| {s_1 s_2 s_3 } \right\rangle \left| 0 \right\rangle  \to \left| {s_1 s_2 s_3 } \right\rangle \left( {\cos \theta _s \left| 0 \right\rangle  + \sin \theta _s \left| 1 \right\rangle } \right),
\end{equation}
where $\cos \theta _s$ is given in \eqref{cosine_rotation} with $ m_s  = J\left( {s_1  + s_2  + s_3 } \right)$. In this case, we have,
\begin{eqnarray}
	W_s &=& \Lambda e^{K\left( {s_1 s_2  + s_2 s_3  + s_1 s_3 } \right)}, \\
	\Lambda  &=& 2\left( {\cosh ^3 \left( J \right)\cosh \left( {3J} \right)} \right)^{1/4}, \nonumber\\
	K &=& \left( {1/4} \right)\ln \left( {\cosh \left( {3J} \right)/\cosh \left( J \right)} \right). \nonumber
\end{eqnarray}
This suggests that to compensate for the renormalization effect of adding the fourth spin, 
we can replace $J$ with $J - K$ in \eqref{renorm_cond_1} and \eqref{Lambda_and_B} when preparing the CETS of the base triangle of the spin-ice tetrahedron that is indicated as red bonds in Fig. \ref{fig:ice}(a).

\emph{Ising spin chains with next-nearest-neighbor interactions and local magnetic fields  ---} As another application of our renormalization method, we consider the application to spin chain with  next-nearest-neighbor interaction, which generalizes the work done in \cite{Lidar97}. Given a chain of $n$ spins with nearest-neighbor interactions ${{J_i}}$, next-nearest-neighbor interactions ${{L_i}}$, and local fields ${{h_i}}$, the Hamiltonian is,
\begin{equation}\label{spin_Ham}
H = \sum\limits_{i = 1}^{n - 1} {{J_i}{s_i}{s_{i + 1}}}  
+ \sum\limits_{i = 1}^{n - 2} {{L_i}{s_i}{s_{i + 2}}}  + \sum\limits_{i = 1}^n {{h_i}{s_i}}. 
\end{equation}
For $i \ge 3$, we define a 2-renormalizable controlled-rotation on the $i$-th qubit:
\begin{equation}\label{2_c_rotation_with_field}
\left| {{s_{i - 2}}{s_{i - 1}}} \right\rangle \left| 0 \right\rangle  \to \left| {{s_{i - 2}}{s_{i - 1}}} \right\rangle \left( {\cos {\theta _i}\left| 0 \right\rangle  + \sin {\theta _i}\left| 1 \right\rangle } \right),
\end{equation}
with $m_s = {L_{i - 2}}{s_{i - 2}} + {J_{i - 1}}{s_{i - 1}} + {h_i}{s_i}$ in \eqref{cosine_rotation}
determining the rotation angles. Using 
\begin{equation}
g_i^{ \pm  \pm } \equiv 2 \cosh \left( {{h_i} \pm {J_{i - 1}} \pm {L_{i - 2}}} \right),
\end{equation}
and following the procedure in \eqref{defineWs}, we can rewrite
\begin{equation}\label{renorm_cond_2}
e^{- \beta m_s} + e^{\beta m_s} = 
{\Lambda _i}{e^{{ \beta B_{i}}{s_{i - 2}}{s_{i - 1}}}}{e^{{\beta C_i}{s_{i - 1}}}}{e^{{\beta D_i}{s_{i - 2}}}}.
\end{equation}
This equality is satisfied for the following choices 
\begin{eqnarray}\label{couplings}
    {\Lambda _i} &=& {\left( {g_i^{ +  + }g_i^{ +  - }g_i^{ -  + }g_i^{ -  - }} \right)^{1/4}}, \\
	{B_i} &=& ( {1/4\beta} )\ln( {g_i^{ +  + }g_i^{ -  - }/g_i^{ -  + }g_i^{ +  - }} ),  \nonumber \\
	{C_i} &=& ( {1/4\beta} )\ln ( {g_i^{ +  - }g_i^{ +  + }/g_i^{ -  - }g_i^{ -  + }} ),  \nonumber \\
	{D_i} &=& \left( {1/4\beta} \right)\ln \left( {g_i^{ -  + }g_i^{ +  + }/g_i^{ -  - }g_i^{ +  - }} \right). \nonumber
\end{eqnarray}
After the controlled rotation on the $i$-th qubit, the spin-spin coupling between the 
spins $i-2,i-1$ is thus renormalized as ${J_{i - 2}} \to {J_{i - 2}} + {B_i}$. 
Also, the local magnetic fields for these spins are renormalized as ${h_{i - 1}} \to {h_{i - 1}} + {C_i}$ 
and ${h_{i - 2}} \to {h_{i - 2}} + {D_i}$. 
Note that the next-nearest-neighbor coupling ${L_{i - 2}}$ term remains unchanged.

Our goal is a CETS of the Hamiltonian \eqref{spin_Ham}.
We can eliminate the unwanted renormalization effects with the following procedure: 
(i) for $\tt i=1..n$, initialize $ {\tt h[ i ]} = {h_i}$, ${ \tt J [ i ]} = {J_i}$ and ${\tt L [ i]} = {L_i}$; (ii) compute $B_i$, $C_i$ and $D_i$ from \eqref{couplings} using the values stored in $( \tt { h[ i ], J [ i ], L [ i ]} )$; put $ {\tt h[ {i - 1} ]} = {\tt h[{i - 1]}} - {C_i}$, ${\tt h[ {i - 2} ]} = {\tt h[{i - 2}]} - {D_i}$ and ${\tt J [ {i - 2}]} = {\tt J[{i - 2}]} - {B_i}$; let $\tt i=i-1$. (iii) Repeat step (ii) until $\tt i=3$. The
desired CETS can then be obtained by a sequence of 2-locally controlled rotations using the stored values in $( \tt { h[ i ], J [ i ], L [ i ]} )$ instead of the original $h_i, J_i$ and $L_i$.

 
\emph{Conclusion ---} To summarize, we have developed an algorithm which identifies a class of classical spin problems that can be simulated efficiently with a quantum computer. In this class of problems, our method scales efficiently compared with MCMC methods, as it is independent of the gap of the Markov chain and temperature. On the other hand, we believe that the tools developed here could be useful for classifying the complexity classes of certain spin models. An avenue for further research is the complexity classification of spin systems by their $t$-renormalizability, which may suggest a deeper understanding of the connection between complexity theory and quantum simulation.



\begin{acknowledgments}
We are grateful to the following funding sources: NSF grant EIA-01-21568 and Croucher Foundation for M.H.Y; European Project OP CE QUTE ITMS NFP 26240120009 and Slovak Research and Development Agency contract APVV LPP-0430-09 for D.N; DARPA under the Young Faculty Award N66001-09-1-2101-DOD35CAP, the Camille and Henry Dreyfus Foundation, and the Sloan Foundation for A.A.G;  Army Research Office under Contract No. W911NF-07-1-0304 for A.A.G and J.D.W.
\end{acknowledgments}

\bibliographystyle{apsrev}
\bibliography{thermal}

\end{document}